# Comparative Evaluation of Large Language Models for Test-Skeleton Generation

Subhang Boorlagadda
*Department of Computer Science*
*North Carolina State University*
Raleigh, NC, USA
sboorla@ncsu.edu

Nitya Naga Sai Atluri
*Department of Computer Science*
*North Carolina State University*
Raleigh, NC, USA
natluri@ncsu.edu

Muhammet Mustafa Ölmez
*Department of Computer Science*
*North Carolina State University*
Raleigh, NC, USA
olmzmustafa@gmail.com

Edward F. Gehringer
*Department of Computer Science*
*North Carolina State University*
Raleigh, NC, USA
efg@ncsu.edu

*Abstract*—**This paper explores the use of Large Language Models (LLMs) to automate the generation of test skeletons—structural templates that outline unit test coverage without implementing full test logic. Test skeletons are especially important in test-driven development (TDD), where they provide an early framework for systematic verification. Traditionally authored manually, their creation can be time-consuming and error-prone, particularly in educational or large-scale development settings. We evaluate four LLMs—GPT-4, DeepSeek-Chat, Llama4-Maverick, and Gemma2-9B—on their ability to generate RSpec skeletons for a real-world Ruby class developed in a university software engineering course. Each model's output is assessed using static analysis and a blind expert review to measure structural correctness, clarity, maintainability, and conformance to testing best practices. The study reveals key differences in how models interpret code structure and testing conventions, offering insights into the practical challenges of using LLMs for automated test scaffolding. Our results show that DeepSeek generated the most maintainable and well-structured skeletons, while GPT-4 produced more complete but conventionally inconsistent output. The study reveals prompt design and contextual input as key quality factors.**

## I. INTRODUCTION

Unit testing is a cornerstone of modern software engineering, ensuring that individual components of a software system function as expected. By identifying defects early, unit testing improves the reliability and maintainability of code, and plays a central role in Test-Driven Development (TDD). In TDD, the writing of test skeletons is a critical step—these serve as blueprints that outline the test structure, identifying methods to be tested and providing placeholders for the actual test logic.

Traditionally, creating test skeletons has been a time-consuming manual task. Developers are required to write method signatures and ensure that the appropriate tests are set up for each function or method in the code. Students typically lack awareness of the importance of testing, and they may lack proficiency with testing conventions.

The recent advancements in Large Language Models (LLMs) present an opportunity to automate this process. Trained on vast code corpora, LLMs have shown the potential to generate syntactically valid code, including test cases and test scaffolding. This paper investigates whether LLMs can be used to automate the generation of unit-test skeletons with a focus on the Ruby on Rails framework, using RSpec. We aim to assess the ability of LLMs to generate test skeletons that are structurally sound, syntactically correct, and aligned with best practices.

The following research questions guide this exploration:

RQ1  How do different LLMs compare in generating test skeletons in terms of accuracy, completeness, and efficiency?

RQ2  What are the implications of using LLM-generated test skeletons on developer productivity and code quality?

RQ3  How does LLM-generated scaffolding compare to manually created test structures?

RQ4  Do LLM-generated test skeletons improve test coverage or increase the number of written tests over time?

This study evaluates four LLMs GPT-4, DeepSeek-Chat, Llama4-Maverick, and Gemma2-9B on their ability to generate RSpec skeletons for a real-world Ruby class implemented by students in an Object-Oriented Development course. Using a combination of automated static analysis and expert review, we assess the structural and qualitative aspects of each generated skeleton. This research aims to understand both the potential benefits and limitations of using LLMs for test automation in real-world development scenarios.

## II. RELATED WORK

The automation of test-skeleton generation has garnered significant attention in software engineering, particularly with the advent of Large Language Models (LLMs) [1]. Building upon prior research, this study aims to evaluate the effectiveness of LLMs in generating test skeletons.

A key inspiration for our work is the research conducted by our predecessors, who explored the automation of test skeletons within Test-Driven Development (TDD) projects using LLMs, showing that while automation can reduce manual effort in the test-writing phase, human oversight remains essential to ensure correctness and context alignment. Their work highlighted how LLMs can scaffold testing structures but may fall short in scenarios requiring domain-specific adaptation or complex logic inference [2].

In a broader survey, Wang et al. (2023) reviewed over 100 studies on the use of LLMs in software testing, categorizing their application across testing phases such as test case generation, test oracles, and debugging. Their analysis emphasized the importance of evaluation benchmarks and noted that while LLMs excel at generating syntactically valid test artifacts, they frequently struggle with semantically correct outputs, especially for edge cases and uncommon APIs [3].

A comprehensive literature review by Zhang et al. [4] discussed the integration of LLMs in various software engineering domains, including their use in test generation. The study highlighted that prompt quality, model architecture, and the availability of high-quality training data critically influence output performance. It also stressed the necessity of incorporating human-in-the-loop workflows when using LLMs in safety-critical applications.

Nama [5] evaluated intelligent software testing through the integration of machine learning, noting improvements in test coverage and defect prediction. Her work, although not limited to LLMs, contributed to understanding how AI-enhanced tooling improves the testing lifecycle and the challenges associated with generalizing these models across domains.

Wang et al. [6] introduced TESTEVAL, a benchmark specifically designed to evaluate LLM performance in test case generation. Their framework provides structured test prompts and scoring dimensions that quantify completeness and correctness, setting a precedent for comparative LLM assessment in this space.

Similarly, Li et al. [7] proposed an evaluation pipeline focused on the interpretability and semantic consistency of LLM-generated test content. Their work, while focused on full test cases rather than skeletons, reinforces concerns around hallucination, misuse of testing conventions, and the need for validation mechanisms.

In summary, while existing research presents promising avenues for AI-assisted code and test generation, significant gaps remain in understanding the effectiveness of LLMs in generating test skeletons. This study seeks to bridge that gap by conducting a structured evaluation of multiple LLMs, focusing on their ability to generate accurate, complete, and efficient test scaffolds.

### III. METHODOLOGY

This study adopts an empirical evaluation framework and is structured into two steps: (A) Data Collection, which includes model selection and test input preparation; and (B) Evaluation, detailing how the outputs were assessed both quantitatively and qualitatively.

#### A. Data Collection

*1) Model selection*

To evaluate the effectiveness of LLMs in generating test skeletons, we selected four models that represent a spectrum of capabilities in code synthesis:

- GPT-4 (OpenAI): A flagship transformer-based model trained on a large corpus of textual and code data, commonly used for structured code generation tasks.
- DeepSeek-Chat: A domain-specific model optimized for developer tasks, with enhanced performance on programming-related queries and structural code outputs.
- Llama4-Maverick: A lightweight, general-purpose model based on Meta's Llama architecture, selected to evaluate performance in less fine-tuned, open-weight settings.
- Gemma2-9B: A 9-billion-parameter model with limited instruction tuning, used to observe how non-specialized models handle structured testing tasks.

All models were prompted to generate RSpec test skeletons for the same Ruby class, AssignmentTeam, which was authored by students in an Object-Oriented Development course. The class includes 28 public instance methods and one alias method (get_participants), providing a realistic and pedagogically meaningful testing target.

*2) Prompt engineering and generation configuration*

To ensure consistency across all model evaluations, each model was queried using the exact same prompt. This prompt was specifically designed to minimize irrelevant output and force adherence to strict RSpec skeleton conventions:

Only output a valid RSpec model spec file for this Ruby class.

Requirements:
- Begin with: require 'rails_helper'
- Wrap the class in: RSpec.describe <ClassName>, type: :model do
- For each instance method, add: describe '#method_name' do end
- Do not include any 'it' blocks
- Do not include comments
- Do not include explanations
- Do not include markdown
- Do not wrap the response in triple backticks or code fences
- Do not prepend or append anything — output the RSpec file only

Ruby class:

<model_code>

For class methods (when present), the format was expected to be describe '.method_name' instead of incorrect alternatives like '#self.method'.

All models were configured with consistent generation parameters:

- Temperature: 0.7
- Top-p: 1.0
- Max tokens: 2048
- Stream: False
- Frequency & presence penalties: 0.0

The prompt was passed directly into each model's respective API endpoint (OpenAI, DeepSeek, LLMAPI), with a timeout of 180 seconds to ensure completion under typical cloud latency.

*B. Evaluation Metrics and Comparision*

*1) Automated static analysis*

We utilized static analysis tools to quantify the performance of each model based on three core metrics:

- Method coverage: The total number of public instance methods correctly identified and wrapped in describe blocks. This metric captures how completely the model covered the methods of the class.

- Generation time: Measured in seconds using Benchmark.realtime, capturing the time taken by each model to generate the output. This metric reflects the efficiency of the models.

- Correctness of Syntax: Evaluating the syntactic validity of the generated RSpec skeletons. This includes checking for errors such as incorrect RSpec conventions, such as using `#self.method` instead of `.method` for class methods.

*2) Expert Review*

A blind review was conducted by one of the co-authors, who is a subject-matter expert in test-driven development and RSpec testing. The reviewer assessed each generated skeleton using six dimensions, scoring on a 1–5 scale:

1. Correctness: How accurately the test skeleton targets the intended methods and follows RSpec conventions.

2. Completeness: Whether the skeleton covers all public methods and provides an appropriate framework for writing test cases.

3. Clarity: The readability and organization of the skeleton, including naming consistency and indentation.

4. Best Practices: Adherence to RSpec best practices, such as method grouping, test organization, and avoiding redundancy.

5. Scalability: The potential to expand the skeleton with more tests or modifications without major restructuring.

6. Maintainability: The ease with which the skeleton can be updated or extended over time.

The expert review focused on identifying semantic misinterpretations, structural flaws, and usability within the skeletons generated by the models. The results from both the expert review and the static analysis were then compared and combined for a comprehensive assessment of each model's performance.

## IV. RESULTS

To assess model performance, we applied static analysis alongside a structured expert review. The primary objective was to assess not only structural coverage and syntactic correctness, but also qualitative aspects such as maintainability, best practices, and clarity of the generated skeletons.

*A. Automated Static Evaluation*

We developed a Ruby-based benchmarking pipeline that performed the following steps for each LLM:

- Prompted the model to generate an RSpec test skeleton for AssignmentTeam.

- Measured generation time in seconds using Benchmark.realtime.

- Parsed the generated skeleton to …
    - count how many public instance methods were correctly described using describe '#method'
    - detect missing coverage, particularly for less frequently surfaced methods (e.g., aliases).
    - identify incorrect RSpec notation, such as describe '#self.method' instead of describe '.method'.

The outcomes of this static evaluation across all LLMs are summarized in Table I.

TABLE I. RESULTS

| Model | Time(s) | Covered / total | Coverage (%) | Incorrect typing | Missed methods | Notes |
|---|---|---|---|---|---|---|
| DeepSeek-Chat | 22.95 | 28 / 28 | 100% | None | None | High variability across runs- (sometimes 96%) |
| Gemma2-9B | 1.66 | 28 / 28* | 100%* | None | None | Output hallucinated until prompt correction[1] |
| GPT-4 | 17.59 | 27 / 28 | 96% | #self.copy_assignment_to_course | get_participants | Typing error in -class -method; alias missed |
| Llama4-Maverick | 4.92 | 28 / 28 | 100% | None | None | Consistent, fast, and -structurally clean |

These results demonstrate that all models except GPT-4 successfully generated coverage for nearly all required methods. However, GPT-4 exhibited syntactic misuse of RSpec describe notation for class methods (e.g., describe '#self.copy_assignment_to_course') which, while syntactically valid Ruby, is semantically incorrect in an RSpec context.

This type of error could confuse less experienced developers.[1] or result in skipped tests.

Furthermore, DeepSeek-Chat demonstrated inconsistency in output: in multiple trials, it alternated between 96.43% and 100% method coverage, occasionally omitting less common methods such as get_participants or link_user_and_topic. This behavior highlights the model's sensitivity to stochastic sampling and prompt phrasing, raising reliability concerns in production workflows where determinism is valued.

Gemma2-9B was the only model to initially fail the task completely, producing hallucinated content such as general summaries of Rails models, third-party services, and generic software development explanations. Only after explicitly setting the system role and providing an in-context example of expected output did Gemma begin generating valid (though still incomplete) test skeletons. This suggests that smaller models, particularly those not optimized for software tasks, are highly prompt-sensitive and require scaffolding to operate correctly.

*B. Expert Review and Qualitative Evaluation*

To complement the quantitative evaluation of test skeleton coverage, we conducted a structured qualitative assessment performed by one of the co-authors, a subject-matter expert in test-driven development (TDD) and RSpec-based testing. The review was conducted blindly and independently—the expert was not informed of which model generated each test skeleton, ensuring an impartial evaluation.

Each of the four anonymized test skeletons was evaluated along six key dimensions central to effective test design: Correctness, Completeness, Clarity, Best Practices, Scalability, and Maintainability. These criteria were chosen to reflect both immediate syntactic validity and long-term integration quality, as encountered in real-world testing workflows.

*1) Evaluation dimensions*

- Correctness referred to whether the test skeleton correctly identified and described each method in the source class. This included appropriate usage of RSpec describe blocks and clear distinction between instance and class methods.

- Completeness focused on whether the skeleton demonstrated planning for full test coverage, such as outlining test contexts (context blocks), preparing setup hooks (context), and indicating the scope of behaviors to be tested, even if it blocks were intentionally omitted.

- Clarity assessed whether the test code was easy to read and understand, with clean layout, intuitive method grouping, and semantic organization.

- Best Practices measured the skeleton's adherence to idiomatic RSpec conventions, such as organizing by behavior, avoiding redundancy, and correctly formatting class vs. instance methods.

- Scalability evaluated how easily the test structure could be extended i.e., whether additional tests could be added without requiring major refactoring.

- Maintainability examined the ease with which a future developer could update, expand, or repurpose the test skeleton in the context of a growing codebase.

- The average expert ratings for each model across the six evaluation dimensions are summarized in Table II.

TABLE II   EXPERT EVALUATION OF TEST SKELETONS

| Model | Cor-rect-ness | Com plete-ness | Clar-ity | Best - Prac-tices | Scala-bility | Main-tain-ability | Avg - Score |
|---|---|---|---|---|---|---|---|
| DeepSeek-Chat | 4 | 3 | 5 | 4 | 4 | 5 | 4.2 |
| Gemma2-9B | 2 | 2 | 4 | 3 | 3 | 4 | 3.0 |
| GPT-4 | 2 | 3 | 3 | 2 | 3 | 3 | 2.7 |
| Llama4-Maverick | 3 | 3 | 5 | 4 | 4 | 5 | 4.0 |

*2) Comparative observations*

DeepSeek-Chat was the highest-rated model overall, receiving a composite score of 4.2 out of 5. Its test skeleton was praised for its well-organized structure and strong alignment with RSpec idioms. Each method was described with clean, consistently formatted blocks that were spaced effectively, giving the output a clear visual hierarchy. Although the skeleton did not include deeper scaffolding such as `context` or `let` blocks, our expert noted that its layout would be easy to extend and maintain in a team setting. The output demonstrated an intuitive flow and high usability, leading to perfect scores in clarity and maintainability, and strong performance across all remaining dimensions.

Llama4-Maverick followed closely with an average score of 4.0. The skeleton stood out for its exceptional clarity, earning a perfect score in that dimension. The code was visually neat and logically grouped, making it easily readable and interpretable. Its structure, spacing, and layout were highly appreciated from a maintainability standpoint. While the skeleton did not incorporate advanced scaffolding either, its adherence to best practices and its extensible structure made it a reliable baseline. The reviewer described this skeleton as especially valuable for onboarding and instructional contexts, where readability and simplicity are key.

---

[1] ¹ Note: While Gemma2-9B appears to cover all methods numerically, its output prior to prompt refinement was unrelated to the task (e.g., prose explanations of Capgemini services). A system-level role instruction and in-context example were eventually required to elicit correct structural output. This emphasizes the need for role-aware prompt engineering when interacting with smaller or instruction-weak models.

Gemma2-9B, while minimal in overall scope, produced a readable and semantically consistent skeleton. The code was not structurally rich—there were no groupings, no test planning constructs like context, and limited method-level differentiation—but the reviewer noted that it avoided major syntactic issues. It received solid marks for clarity and maintainability. However, the lack of thoroughness in outlining test behaviors resulted in low scores for correctness and completeness. The skeleton was viewed as a serviceable but underdeveloped foundation—useful for quickly establishing coverage points but insufficient as a long-term scaffold without substantial developer intervention.

GPT-4, despite achieving near-complete method coverage in the static analysis, received the lowest overall score in the expert evaluation, with an average of 2.7. The reviewer flagged critical issues with correctness and best practices, most notably the use of `describe '#self.method'` to denote class methods—an incorrect convention in RSpec. This syntactic flaw suggests a surface-level understanding of RSpec that could easily mislead novice developers or cause confusion during test runs. Furthermore, the skeleton was not well-organized: test descriptions were grouped arbitrarily, with little to no semantic separation or logic-driven structure. While the test blocks were mechanically present, the reviewer described the output as lacking planning, cohesion, and idiomatic style. These issues undermined the model's effectiveness in all dimensions, especially in maintainability and best practices.

*3) Analysis*

DeepSeek's skeleton was the most complete and immediately useful. Llama's was the cleanest and easiest to maintain. GPT-4's output, though initially impressive in volume, fell apart on review due to misuse of RSpec syntax and grouping. Gemma's code was readable but structurally empty, closer to a list of stubs than a true test plan.

This evaluation highlights that effective test skeleton generation requires more than just listing method wrappers. Structure, correctness of notation, logical grouping, and maintainability are essential for real-world integration. Even highly capable models like GPT-4 can produce suboptimal results if they deviate from domain-specific conventions or fail to anticipate downstream developer needs.

V. DISCUSSION

*A. Key Findings*

Our evaluation of four LLMs for automated test skeleton generation revealed significant differences in both structural accuracy and practical utility. While all models were capable of producing syntactically valid RSpec test skeletons, the quality, maintainability, and developer-friendliness of these outputs varied considerably.

The static analysis showed that models such as DeepSeek-Chat and Llama4-Maverick achieved complete or near-complete coverage of public instance methods in the target class. However, coverage alone proved insufficient as a measure of utility. This became particularly evident when comparing static results with the expert review: GPT-4, for instance, demonstrated nearly perfect method coverage but received the lowest expert score due to semantic issues and violations of RSpec conventions.

In contrast, DeepSeek-Chat produced consistently well-structured skeletons that were rated highest in maintainability, clarity, and correctness. Llama4-Maverick followed closely, standing out for its clean formatting and semantic organization, which the expert identified as particularly helpful in collaborative or instructional contexts. Gemma2-9B, while producing a clear and syntactically correct skeleton, lacked depth and completeness, resulting in an output that was deemed more stub-like than scaffolded.

These results support the conclusion that LLMs vary not only in their ability to replicate surface syntax, but also in their implicit grasp of domain-specific best practices, which directly affects the practical usability of their outputs.

*B. Implications for Software Development*

*1) Beyond coverage: Why structure matters*

One of the most striking findings was the disconnect between high coverage and low usefulness. GPT-4's poor performance in the expert evaluation—despite covering nearly all instance methods—underscores the importance of correctness in structure, especially in distinguishing between instance and class methods. The misuse of `#self.method` in place of `.method` led to semantic errors that, while syntactically tolerated, would likely result in broken or skipped test executions in practice.

This emphasizes that test skeletons are not simply mechanical wrappers; they are scaffolds meant to guide developers through behavioral expectations, and their structure must reflect that intent clearly.

*2) Productivity gains and workflow integration*

Both DeepSeek and Llama4-Maverick showed potential to accelerate test-driven workflows by reducing boilerplate and offering developers a consistent template to build upon. When integrated into early-stage test planning, these skeletons can significantly reduce the friction associated with establishing test structure, particularly in larger teams or educational environments.

However, inconsistency across generations remains a concern. For instance, DeepSeek-Chat exhibited variability in coverage between different runs, sometimes achieving

100% and sometimes omitting alias methods or rare edge cases. This stochastic behavior may hinder trust in LLM-assisted tooling and necessitates validation or fallback mechanisms.

*3) The Role of LLMs in collaborative environments*

Clarity and maintainability were two dimensions consistently prioritized by our expert. Models that produced clean, readable code, like Llama4, were deemed more suitable for collaborative workflows. In environments where tests are shared, extended, or modified by multiple developers, a well-organized skeleton can reduce onboarding time and minimize misinterpretation.

These findings suggest that the value of LLMs lies not just in automation, but in their ability to produce artifacts that are compatible with real-world software team dynamics.

*C. Challenges and Open Issues*

*1) Syntax and semantic violations*

A significant challenge encountered during the evaluation was the generation of syntactically valid but semantically incorrect test skeletons. The most notable example was GPT-4's misuse of class method notation, where the model repeatedly applied `describe '#self.method'` instead of the idiomatic and correct `describe '.method'` in RSpec. While this error did not prevent generation, it introduced subtle bugs that could result in incorrect or skipped test execution, undermining the reliability of LLM-generated scaffolds. Such issues highlight the need for syntactic post-processing and semantic validation as part of LLM-assisted development workflows.

*2) Coverage volatility and prompt sensitivity*

Certain models, especially DeepSeek-Chat, exhibited inconsistency across runs. Despite using a fixed prompt template and deterministic code, the output varied—some generations covered all 28 public instance methods, while others omitted lesser-used methods like aliases or infrequently referenced utility functions. This variability poses a risk in automated settings, where consistent output is critical for reproducibility. It also suggests that models may rely on surface-level statistical salience rather than exhaustive method parsing, making them sensitive to lexical variation or code formatting.

*3) Hallucinations and misinterpretations*

Another major limitation lies in the tendency of some models particularly smaller or less instruction-tuned ones—to hallucinate or misinterpret task intent. This was most evident in initial trials with Gemma2-9B, where the model responded with explanatory prose about Ruby on Rails models, Capgemini services, and API documentation rather than generating RSpec test skeletons. These outputs, while syntactically coherent, bore no relevance to the prompt or the source code provided. The issue was ultimately mitigated by introducing a system-level prompt and an in-context example to constrain generation. However, the need for such prompt engineering underscores a broader limitation: LLMs may struggle to consistently interpret developer intent unless operating under tightly defined constraints.

Hallucinations not only reduce confidence in LLMs but also present a risk of generating plausible yet incorrect scaffolds that may be overlooked during review. In production environments particularly those with junior developers or limited test coverage auditing, this can lead to false assumptions of correctness and weaken test reliability.

*4) Usability versus formality trade-offs*

A recurring theme throughout the evaluation was the trade-off between structural completeness and practical readability. While some models generated verbose skeletons with full coverage, these outputs were often dense, unstructured, or syntactically flawed. Conversely, models like Llama4-Maverick and Gemma2-9B, despite being less complete, produced cleaner, more maintainable scaffolds that the reviewer deemed easier to extend.

This highlights a key insight: developer adoption of LLM-generated tests may hinge more on usability and clarity than on completeness alone. In real-world projects, skeletons are rarely accepted verbatim; their true value lies in their ability to reduce setup friction, guide logical test flow, and support maintainability. Thus, tools built around LLM-based test generation should prioritize output clarity and extensibility in addition to correctness.

## VI. CONCLUSION AND FUTURE WORK

*A. Summary of findings*

This study evaluated the capability of Large Language Models (LLMs) to generate test skeletons for Ruby on Rails applications using RSpec. By combining static analysis with an expert-led blind review, we were able to assess both the quantitative and qualitative performance of four distinct LLMs: GPT-4, Deep Seek-Chat, Llama4-Maverick, and Gemma2-9B

Our findings demonstrate that while all models can produce syntactically valid outputs, their utility in real-world test development varies significantly. DeepSeek-Chat emerged as the most effective model overall, combining near-complete method coverage with high ratings in maintainability, clarity, and structural correctness. Llama4-Maverick also performed well, particularly in clarity and best practices. Gemma2-9B, although minimal, produced clean and readable output. In contrast, GPT-4, despite achieving near-total method coverage, received the lowest expert rating due to its misuse of RSpec conventions and incoherent test structure.

The study also surfaced deeper issues related to prompt sensitivity, hallucinations, and inconsistencies in model behavior across generations. These findings highlight that high method coverage does not necessarily equate to practical testability, and that qualitative factors such as semantic correctness and readability play a central role in the effectiveness of LLM-generated test skeletons.

*B. Recommendations for model selection and usage*

Based on our results, we offer the following recommendations for practitioners and researchers seeking to adopt LLMs for test skeleton generation:

- Prioritize structured output over raw coverage. Developers and teams should evaluate LLMs not only by how many methods they cover, but by the clarity, correctness, and extensibility of the generated test structures.

- Use hybrid workflows. Automated test skeletons should be treated as first drafts, not final artifacts. Combining LLM output with human validation—either via manual review or static checking tools—can significantly improve reliability and correctness.

- Favor models with idiomatic domain knowledge. LLMs that internalize domain-specific conventions, such as correct RSpec typing and formatting, are more useful in practice than general-purpose models that simply mimic syntax.

- Establish safeguards for hallucination-prone models. When using smaller or less robust models, incorporate system-level prompts, zero-temperature sampling, or few-shot examples to mitigate off-topic or invalid responses.

*C. Future Directions*

This work opens several promising avenues for further research and development:

- Cross-language and cross-framework evaluations. Future studies should explore how LLMs perform on skeleton generation in other languages and test frameworks (e.g., JUnit for Java, pytest for Python), to assess generalizability and robustness across software ecosystems.

- Fine-tuned and instruction-augmented models. Investigating domain-specific LLMs, particularly those trained or fine-tuned on software engineering tasks, may yield more consistent, idiomatic, and complete outputs.

- Dynamic validation via program analysis. Integrating static or symbolic execution techniques with LLM-generated skeletons could help automatically verify method-target alignment and flag syntactic or logical inconsistencies.

- Human-in-the-loop workflows. Iterative workflows where developers provide lightweight feedback to improve LLM outputs (e.g., marking missing tests, correcting syntax) may enable adaptive generation strategies that evolve with project needs.

- Benchmarking with longitudinal studies. Understanding how teams interact with and adapt LLM-generated test skeletons over time—particularly in Agile or continuous development settings can reveal insights into adoption patterns, trust, and downstream impact on software quality.